\documentclass[11pt]{article}
\usepackage{epsfig}
\usepackage{amsmath}
\usepackage{amssymb}
\usepackage{mathrsfs}
\usepackage{graphicx}
\newsavebox{\PSLASH}
\sbox{\PSLASH}{$p$\hspace{-1.8mm}/}

\newsavebox{\KSLASH}
\sbox{\KSLASH}{$k$\hspace{-1.8mm}/}

\begin{document}

\vspace{4cm}
\begin{center}
{\Large\bf{Top Quark Asymmetries and Unparticle Physics at the Tevatron and LHC }}\\

\vspace{1cm}
{\bf Sara Khatibi and Mojtaba Mohammadi Najafabadi }\\
\vspace{0.5cm}
{\sl  School of Particles and Accelerators, \\
Institute for Research in Fundamental Sciences (IPM) \\
P.O. Box 19395-5531, Tehran, Iran}\\

\vspace{2cm}
 \textbf{Abstract}\\
 \end{center}
Among different measured observables of top-antitop quark pairs at
hadron colliders, the forward-backward asymmetry ($A_{\text{FB}}$)
measured by the CDF and D0 collaborations has inconsistency with the
Standard Model prediction. The measured forward-backward
asymmetry grows with $t\bar{t}$ invariant mass.
 Several new physics models have been
proposed to explain this deviation.
We consider the consistency of the parameter space of vector
unparticle (in Flavor-Conserving scenario) with the existing $t\bar{t}$ production measurements.
In particular, we look at the total cross sections at the LHC and Tevatron, differential
cross section with $t\bar{t}$ invariant mass, and the LHC charge asymmetry to identify the
regions in parameter space that can give the desired top
$A_{\text{FB}}$ observed by
the Tevatron.
We show that in spite of the intrinsic tension between the LHC charge
asymmetry and $A_{\text{FB}}$, there exists a region in the unparticle parameters space
where the top $A_{\text{FB}}$ and the LHC charge asymmetry are satisfied
simultaneously. Finally, we show that the consistent region with
$t\bar{t}$ observables is consistent with
the constraints coming from the dijet resonance searches.
\\


\newpage

\section{Introduction}

Top quark, with its mass near to the scale of electroweak symmetry breaking can be more sensitive to new physics at TeV scale than the other
Standard Model (SM) partciels. Most of
its properties have been examined at the Tevatron and LHC and found to be in
agreement with the SM predictions \cite{frank},\cite{werner}, except the observed forward-backward asymmetry in top quark pair production
$(A_{\text{FB}})$ which has about $2 \sigma$ deviation from the SM
expectation. The forward-backward asymmetry ($A_{FB}$)
is defined as the difference between the number of top quark in
the froward ($\cos\theta>0$) and backward ($\cos\theta<0$) region of the detector:
\begin{eqnarray}\label{afb11}
A_{FB} = \frac{N_{t}(\cos\theta > 0)-N_{t}(\cos\theta < 0)}{N_{t}(\cos\theta > 0)+N_{t}(\cos\theta < 0)}
\end{eqnarray}
Where $\theta$ is the top quark production angle in the $t\bar{t}$
rest frame. The SM prediction for $A_{\text{FB}}$ at loop level
is 0.089 \cite{afbsm1}, \cite{afbsm2},\cite{afbsm3},\cite{afbsm4}.
While the recent measurements reported by CDF and D0 are $A_{FB} = 0.158 \pm 0.075$
\cite{cdfafb},\cite{cdfnote}, $A_{\text{FB}} = 0.196 \pm 0.065$ \cite{d0afb}.
We note that the observed forward-backward asymmetry increases with the $t\bar{t}$
invariant mass such that it approaches $0.3$ for $m_{t\bar{t}} \geq$
700 GeV.

Unlike the top $A_{\text{FB}}$, the total $t\bar{t}$ cross section which has been measured in Tevatron is in agreement with the SM prediction \cite{cstev}.
The $t\bar{t}$ differential cross section with the $t\bar{t}$
invariant mass ($d\sigma/dm_{t\bar{t}}$) has been also measured by the
CDF collaboration. The $t\bar{t}$ spectrum has been found to be
consistent with the SM expectation including NLO+NNLL QCD predictions \cite{cdfmtt}.
The measured top pair cross section at the LHC confirms the SM expectation at the NNLO QCD prediction \cite{cslhc}.
The present measured differential cross
section ($d\sigma/dm_{t\bar{t}}$) by the LHC experiments are limited by statistical and
systematic uncertainties \cite{mttlhc}.

It is interesting to note that the $A_{\text{FB}}$ vanishes at the LHC because of the symmetric initial state.
However, another asymmetry at the LHC ($A_{\text{C}}$) can be defined as
the relative difference between top pair events with $|y_{t}| > |y_{\bar{t}}|$ and the events with $|y_{t}| < |y_{\bar{t}}|$:
\begin{eqnarray}
A_{C} = \frac{N_{t}(|y_{t}| > |y_{\bar{t}}|)-N_{t}(|y_{t}|<|y_{\bar{t}})}{N_{t}(|y_{t}| > |y_{\bar{t}}|)+N_{t}(|y_{t}|<|y_{\bar{t}})}
\end{eqnarray}

At the LHC the top quarks produced in the quark-antiquark annihilation process are
statistically more boosted to the beam direction in comparison with the antitop quark. This is because of the fact that
the top quark prefers to fly in the direction of the incident quark which carries a larger longitudinal
momentum. As a consequence, a charge asymmetry as described above is generated.
The ATLAS and CMS measurements for the charge asymmetry are: $A_{\text{C}} = -0.018 \pm 0.036$ \cite{atlasac}, $A_{\text{C}} = 0.004 \pm 0.015$
\cite{cmsac}, and the SM prediction is $A_{\text{C}} = 0.0115$ \cite{afbsm4}.
Within the uncertainties the standard model prediction is in agreement
with the measured values by the LHC experiments.
The charge asymmetry has been measured in various $m_{t\bar{t}}$ bins
by ATLAS and CMS experiments but with large uncertainties therefore, we
use the inclusive measured charge asymmetry in our analysis.

It is notable that some of the SM extensions proposed to explain the Tevatron $A_{\text{FB}}$ also predict sizeable charge asymmetry at the LHC \cite{wp1},\cite{yang},\cite{juan1},\cite{juan2}.
Therefore, in those models there exists a tension between the top forward-backward asymmetry at Tevatron and the LHC charge asymmetry.
From another side, the LHC charge asymmetry measurement is consistent
with the SM expectation consequently the models which predict also
enhancement in $A_{\text{C}}$ are disfavored.
For example, it has been shown that in the $W'$ and $Z'$ models there is a tight correlation between $A_{\text{FB}}$ and
$A_{\text{C}}$. Therefore, these models are not able to explain the
charge asymmetry and $A_{FB}$ at the same time \cite{juan1},\cite{Fajfer:2012si},\cite{khatibi},\cite{ayazi}.
In \cite{efflag}, the effective Lagrangian approach has been utilized to explain the $A_{\text{FB}}$. In this approach an enhancement in
$A_{\text{C}}$ is also expected, in particular at large $t\bar{t}$ invariant mass region. It has been shown in \cite{Fajfer:2012si} that
there is an apparent tension between the forward-backward asymmetry
and the charge asymmetry in axigluon model but there exists an allowed
region compatible with both $A_{\text{FB}}$ and $A_{\text{C}}$.

It seems difficult to develop a model that can produce large $A_{\text{FB}}$ deviated from the SM prediction according to Tevatron measurement but $A_{\text{C}}$
is consistent with the SM value. There are studies on this issue which for example can be found in \cite{juan1},\cite{juan2}.


In this work, we study the effects of color singlet vector unparticles \cite{Georgi:2007ek}, \cite{Georgi:2007si}
on the forward-backward asymmetry and charge asymmetry at Tevatron and LHC, respectively.
We investigate the tension between $A_{\text{C}}$ and $A_{\text{FB}}$ and perform a full
scan on the main unparticle parameters space. In constraining the
unparticle parameters we combine $A_{\text{FB}}$ ($m_{t\bar{t}}$ dependent),
$\sigma_{\text{LHC}}$, and
$\sigma_{\text{TeV}}$ into a global $\chi^{2}$ fit to
obtain $68\%$ C.L. region.
We also require that the resulting region to be consistent with
the constraints coming from the dijet resonance searches.
The organization of this letter is as follows. Next section is devoted to unparticle physics and its effect on top prodution rate. In section
3, we show our numerical calculations and discuss the results. Finally, conclusions are presented in section 4.

\section{Influence of unparticle on top pair production }

The effects of unparticle on top properties at hadron colliders have been intensively studied in the literatures
\cite{Alan:2007ss},\cite{Choudhury:2007cq},
\cite{Arai:2009cp}, \cite{Alan:2007ui}, \cite{Li:2008ik}, \cite{Sahin:2008ty}, \cite{Aliev:2008ng}, \cite{Aliev:2009jm}. Also, there are some
papers in which the top $A_{\text{FB}}$ at the Tevatron has been studied.  In \cite{Chen:2010hm}, the authors have found
the regions of parameters where colored vector unparticle can produce the values of top $A_{\text{FB}}$ and the top pair cross sections
compatible with the Tevatron measurements.
In \cite{Dahiya:2012ka}, the influence of vector and tensor unparticle, including color, on top pair cross section and the forward-backward asymmetry has
been investigated.
However, in these studies the impact of  unparticle on
the LHC charge asymmetry and any possible tension with $A_{FB}$ have not been investigated.

Effective interaction of vector unparticle with SM fields are given as follows \cite{unparticle}:
\begin{eqnarray}
\lambda_1 \frac{1}{\Lambda^{d_\mathcal{U}- 1}}c_v\bar f \gamma_\mu
f O_\mathcal{U}^\mu ~~,~~ \lambda_1 \frac{1}{\Lambda^{d_\mathcal{U}- 1}}c_a\bar f \gamma_\mu \gamma_5
f O_\mathcal{U}^\mu
\end{eqnarray}
Where $\lambda_{1}$ is dimensionless effective couplings labeling vector
unparticle operator. The coefficients $c_{v} , c_{a}$ represent vector and axial vector couplings of vector
unparticle, respectively. The parameter $d_{\mathcal{U}}$ is the scaling dimension of the unparticle operators and $\Lambda$
denotes the effective mass scale above which unparticle is formed. \\
Within the SM at hadron colliders, $t \bar t$ pairs are produced
either via quark-antiquark annihilation or through
 gluon-gluon fusion. With considering new interactions of
vector unparticle with SM fields, only the partonic cross section for $t \bar{t}$ production via quark-antiquark annihilation is
modified, because vector unparticle only interacts with fermionic fields and it does not couple to gluons.
The parton level differential cross section for the process of $q\bar{q}\rightarrow t\bar{t}$
at leading order in the presence of color singlet vector unparticle is as follows \cite{Alan:2007ss}:
\begin{eqnarray}\label{cs}
\frac{d\hat\sigma}{d\hat t}(q\bar q\rightarrow t\bar
t)&=&\frac{A_V^2}{8\pi\hat s^2(\hat
s)^{4-2d_{\mathcal{U}}}}\Big[c_a^4(2m^4-4(\hat s+\hat t)m^2+(\hat
s+\hat t)^2+\hat t^2)\nonumber\\&+&c_v^4(2m^4-4\hat tm^2+(\hat
s+\hat t)^2+\hat t^2)\nonumber\\&+&2c_v^2c_a^2(2m^4-2(3\hat s+2\hat
t)m^2+3\hat
s^2+2\hat t^2+6\hat s\hat t)\Big]\nonumber\\
&+&\frac{d\sigma_{q\bar q}^0}{d\hat t},
\end{eqnarray}
where
\begin{eqnarray}
A_V=\frac{\lambda_1^2A_{d_{\mathcal{U}}}}{2\sin(d_{\mathcal{U}}\pi)\Lambda^{2(d_{\mathcal{U}}-1)}},~ 
   A_{d_\mathcal{U}}=\frac{16\pi^{2}\sqrt{\pi}}{(2\pi)^{2{d_U}}}
     \frac{\Gamma(d_\mathcal{U}+\frac{1}{2})}{\Gamma(d_{\mathcal{U}}-1)\Gamma(2{d_\mathcal{U}})}.
\end{eqnarray}

In the cross sections relation, $\frac{d\sigma_{q\bar q}^0}{d\hat t}$ is the SM contribution.

In Eq.\ref{cs}, for the case that $c_{v} = 1$ is corresponding to vector unparticle, $c_{v} = c_{a} = 1$ is corresponding to
vector unparticle with right-handed coupling to the SM fields and $c_{v} = -c_{a} = 1$ presents the vector unparticle
with left-handed coupling. According to Eq. \ref{cs}, the cross section is similar in both cases with $c_{v} = c_{a} = 1$ and
$c_{v} = -c_{a} = 1$. Therefore, the $t\bar{t}$ cross section and forward-backward asymmetry in this scenario
are chirality independent or blind to left-hand or right-handed couplings.

\section{Numerical Results and Discussion}

In the numerical calculations, the top quark mass has been set $m_{t} = 173$ GeV.
All cross sections at the partonic level is calculated by employing
CTEQ6 parton distribution functions \cite{cteq}.
The calculation is performed at fixed renormalization and factorization scale $\mu_{R} = \mu_{F} = m_{t}$.
We present our numerical result at Tevatron with $\sqrt{s}=1.96 ~\rm TeV$ and at LHC with $\sqrt{s}=7 ~\rm TeV$.
Indeed, the cross sections that we get
from the calculations are the leading order values. Therefore, we scale the tree-level calculation by a $k$-factor of
1.3, so that the leading order calculations match with the higher
order results for the case of $m_{t}=173$ GeV/c$^{2}$. 
This $k$-factor is introduced so that the tree level SM result after
applying $k$-factor gives tha SM higher order results.
The NNLO
cross section of top pair production at Tevatron is 7.08 pb and 163 pb at the LHC with the center-of-mass energy of 7 TeV \cite{kidonakis}.

As we mentioned before, the results are chirality independent and the
right-handed and
left-handed unparticle couplings to the SM fields
give similar cross sections and asymmetries in top pair events.
In the case of having pure vector unparticle i.e. $c_{v} = 1$ and
$c_{a} = 0$, we saw that negligible forward-backward
asymmetry is produced which can not compensate the observed value by Tevatron experiments.

First, we present asymmetries in terms of $d_{\mathcal{U}}$ for three various values of $\Lambda$,
and consider $\lambda_{1}=1$ and $c_{a}=c_{v}=1$. Then we identify an
allowed region in the $d_{\mathcal{U}}$, $\Lambda$ plane
by combining $A_{FB}$ (taking into account data in various $t\bar{t}$
invariant mass bins),
charge asymmetry ($A_{\text{C}}$), $\sigma_{\text{LHC}}$, and
$\sigma_{\text{TeV}}$ into a global $\chi^{2}$ fit.
We concentrate on the values of unparticle parameters which are
physically interesting,
i.e. $1 < d_{\mathcal{U}} < 2$ and $\Lambda$ at the order of few TeV \cite{unitarity}.

\begin{figure}
\centering
  \includegraphics[width=7cm,height=5cm]{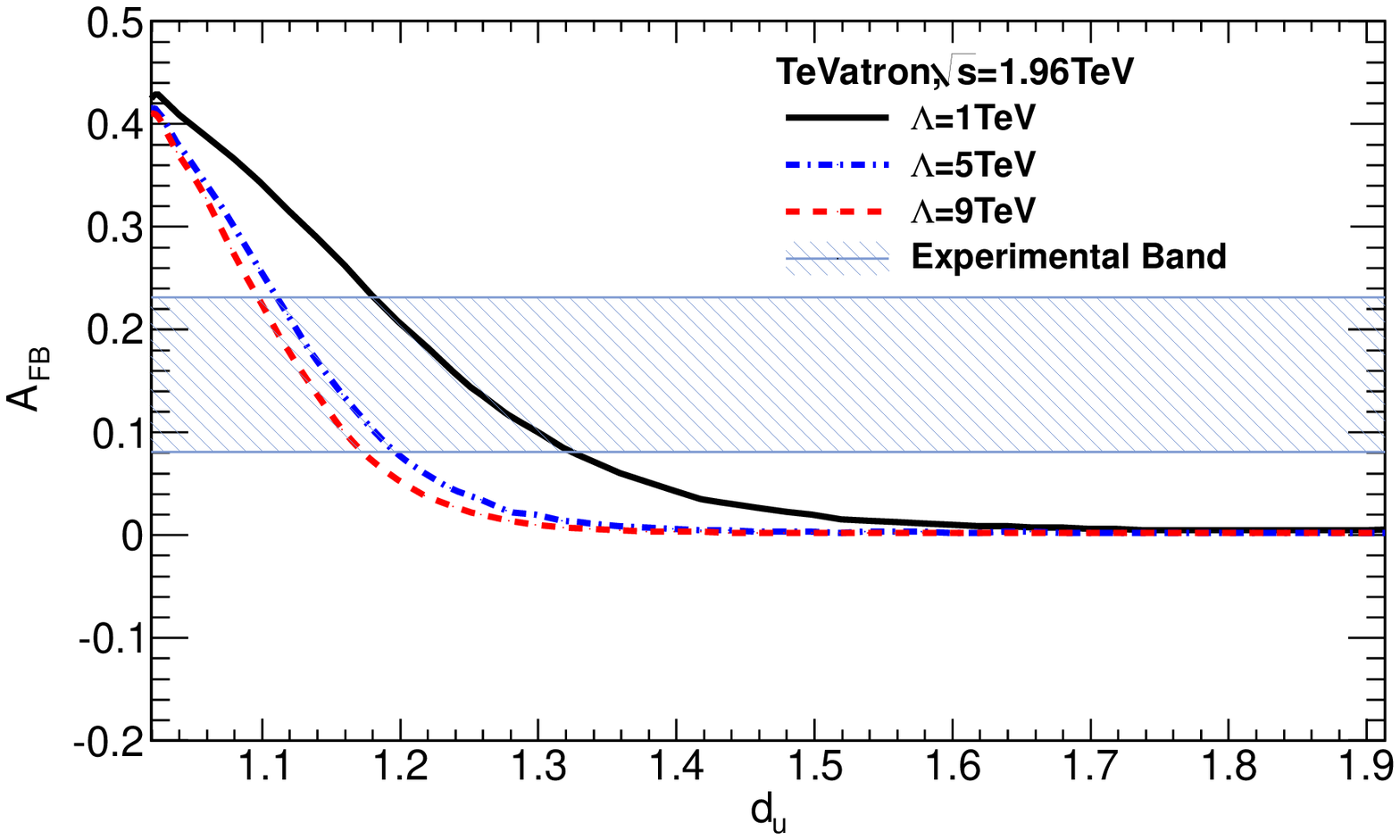}
  \includegraphics[width=7cm,height=5cm]{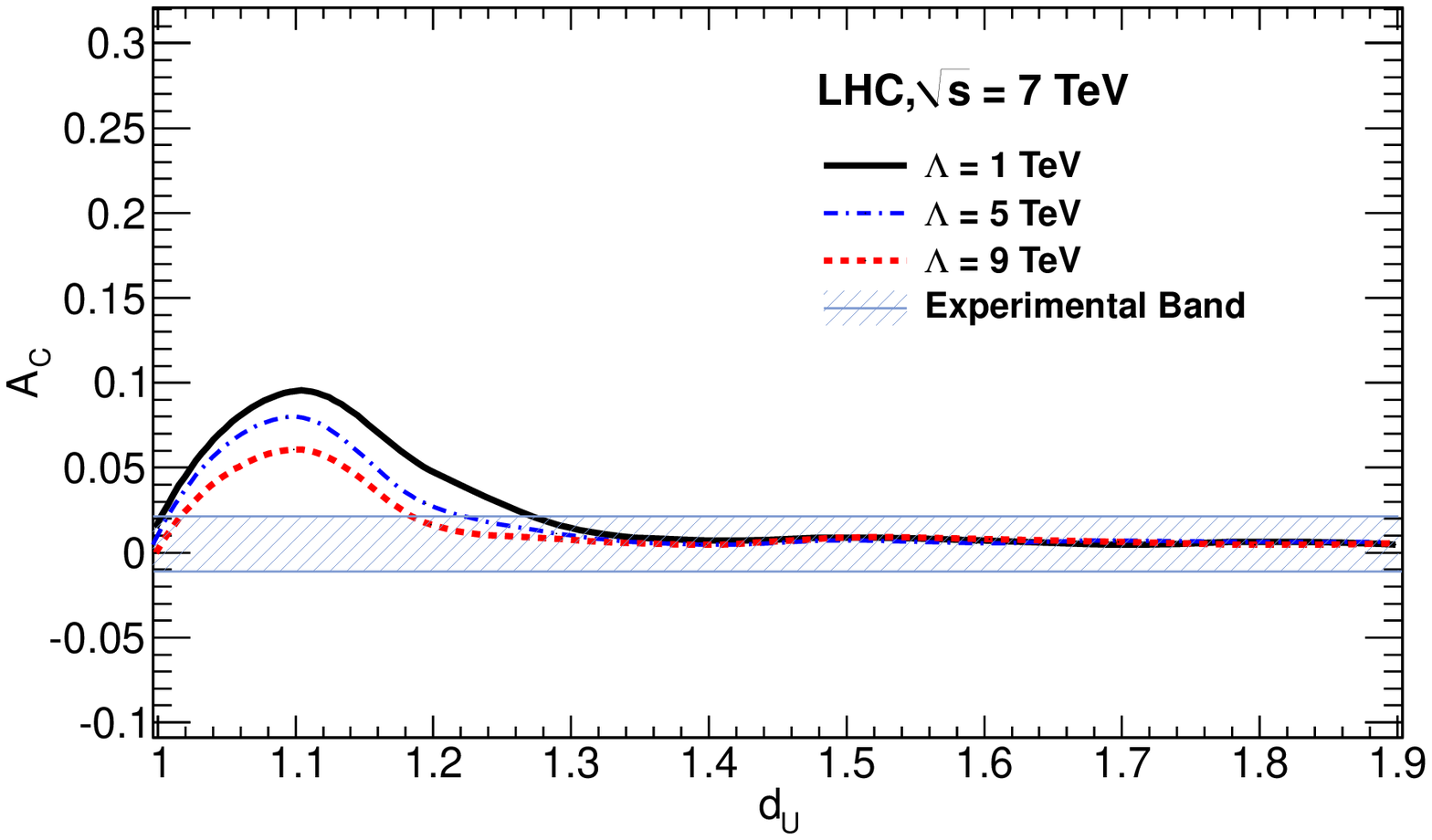}
  \caption{\small{ Left: The forward-backward asymmetry $(A_{\text{FB}})$ in top pair production generated by vector unparticle for $\lambda_{1}=1$ and
  $c_{a}=c_{v}=1$ and various values of $\Lambda$. Shaded region is the band measured by the CDF experiment.
Right: The charge asymmetry at the LHC
with the same parameters as $A_{\text{FB}}$ plot and different values of $\Lambda$.}}\label{asymmetry}
\end{figure}

The forward-backward asymmetry $(A_{\text{FB}})$ at Tevatron and the charge asymmetry at the LHC are shown in Fig. \ref{asymmetry}.
The shaded area is according to the present experimental measurement.
As it can be seen, for a specific value of $d_{\mathcal{U}}$ the forward-backward asymmetry grows when $\Lambda$ decreases, i.e. unparticle can
produce larger asymmetry by assuming small values of $\Lambda$.
Note that for larger values of $\Lambda$, the allowed interval of $d_{\mathcal{U}}$ parameter that
can produce desirable forward-backward asymmetry becomes smaller. According to Fig. \ref{asymmetry} at $\Lambda=1$ TeV, unparticle
with any value of $d_{\mathcal{U}}$ in the range of 1.2 to 1.32 can generate the desired $A_{\text{FB}}$.

The charge asymmetry $A_{\text{C}}$ increases with
increasing $d_{\mathcal{U}}$, reaches to a maximum value at $d_{\mathcal{U}} = 1.1$ then it decreases and
tends to the SM expectation at the tail of $d_{\mathcal{U}}$. The peak position does not
move for various values of $\Lambda$.
The shaded region is according to the CMS measurement. For example, when $\Lambda = 1$ TeV,
unparticle with $d_{\mathcal{U}} \leq 1.28$ is excluded. For larger values of $\Lambda$, the exclusions interval is smaller.

\begin{figure}
\centering
  \includegraphics[width=10cm,height=8cm]{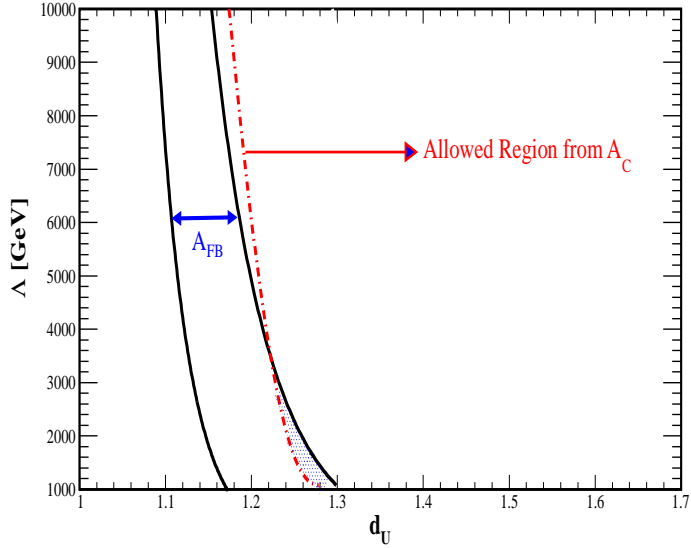}
  \caption{\small{Region of $\Lambda$ (in GeV) in terms of $d_{\mathcal{U}}$ consistent with Tevatron measurements of
the  $t\bar{t}$ forward-backward asymmetry  (region between two solid
black curves).
The consistent region with the LHC charge asymmetry is the area in the
right side of the dotted-dashed red curve. }}\label{combine1}
\end{figure}

In Fig.\ref{combine1}, we present the allowed regions in the plane of $(d_{\mathcal{U}},\Lambda)$ which satisfy
the measured forward-backward asymmetry by Tevatron and the LHC charge asymmetry.
The combination of limits from $A_{\text{C}}$ and the allowed band for
$A_{\text{FB}}$ leads to a very small allowed interval of 1.27 to 1.3
for $d_{\mathcal{U}}$ at the value of $\Lambda = 1$ TeV. 
As it can be seen from Fig.\ref{combine1}, charge asymmetry excludes a
large part of the parameter spaces which could explain the Tevatron
forward-backward asymmetry. For any value of $\Lambda$ above
3400 GeV, the LHC charge asymmetry excludes the points in
$(d_{\mathcal{U}},\Lambda)$ which are consistent with the measured
forward-backward asymmetry.
According to Fig.\ref{combine1}, there is an apparent tension between
the forward-backward asymmetry and charge asymmetry for this model. We
note that this tension gets tighter for large values of $\Lambda$.

It has been shown that there is an intrinsic tension between the
observed large positive forward-backward asymmetry by Tevatron
and the LHC measurement of charge asymmetry \cite{Fajfer:2012si}.
The relation between the Tevatron $A_{\text{FB}}$ and the LHC charge asymmetry $A_{\text{C}}$
is model dependent. Models like $W'$,$Z'$ can generate the desired forward-backward asymmetry but
the LHC charge asymmetry disfavors the regions where $A_{\text{FB}}$ is
generated according to the Tevatron measurements.
In contrary, there can be found models such as axigluon which
can produce all related observables according to the Tevatron and LHC measurements.

\begin{figure}
\centering
    \includegraphics[width=7cm,height=5cm]{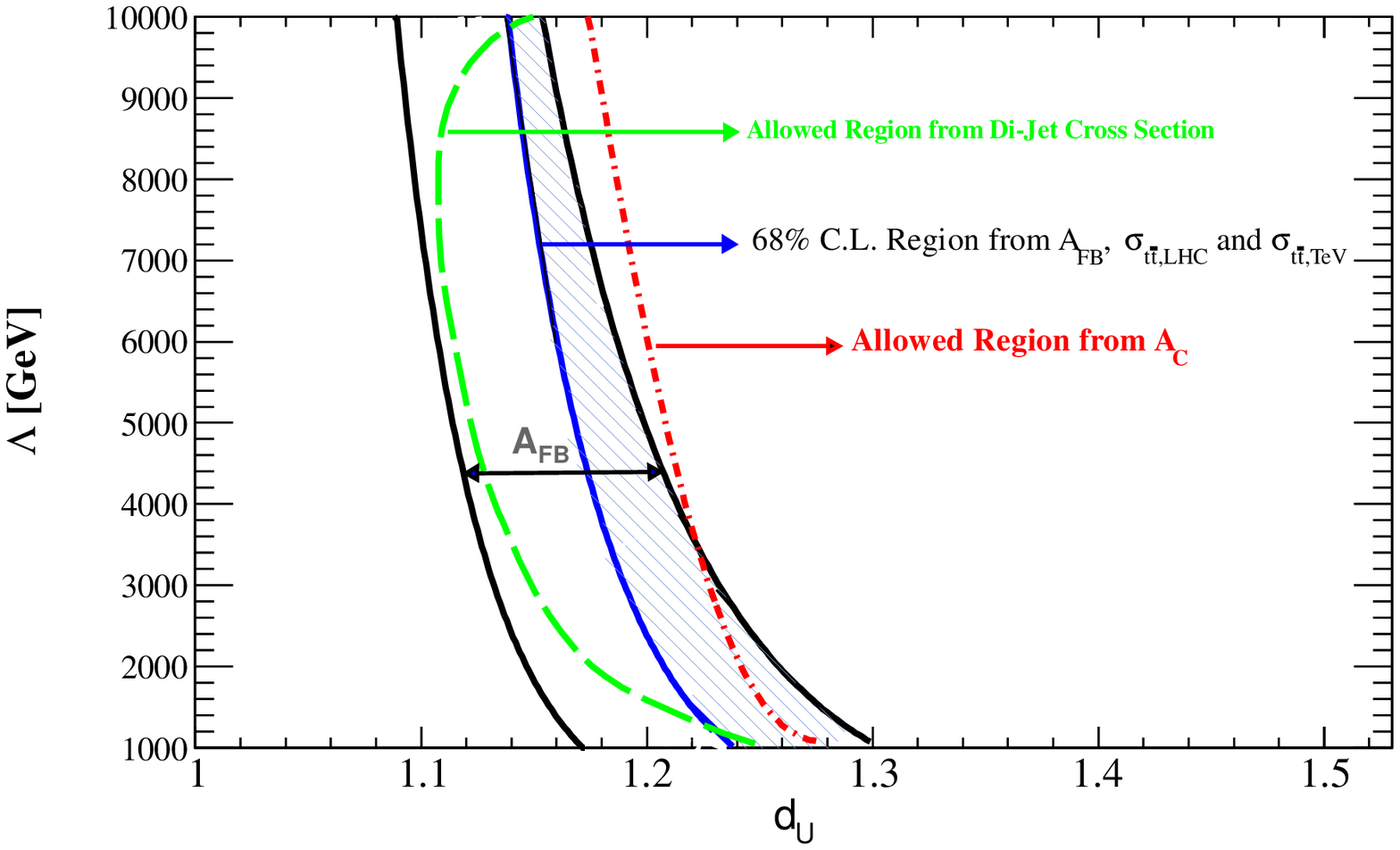}
    \includegraphics[width=7cm,height=5cm]{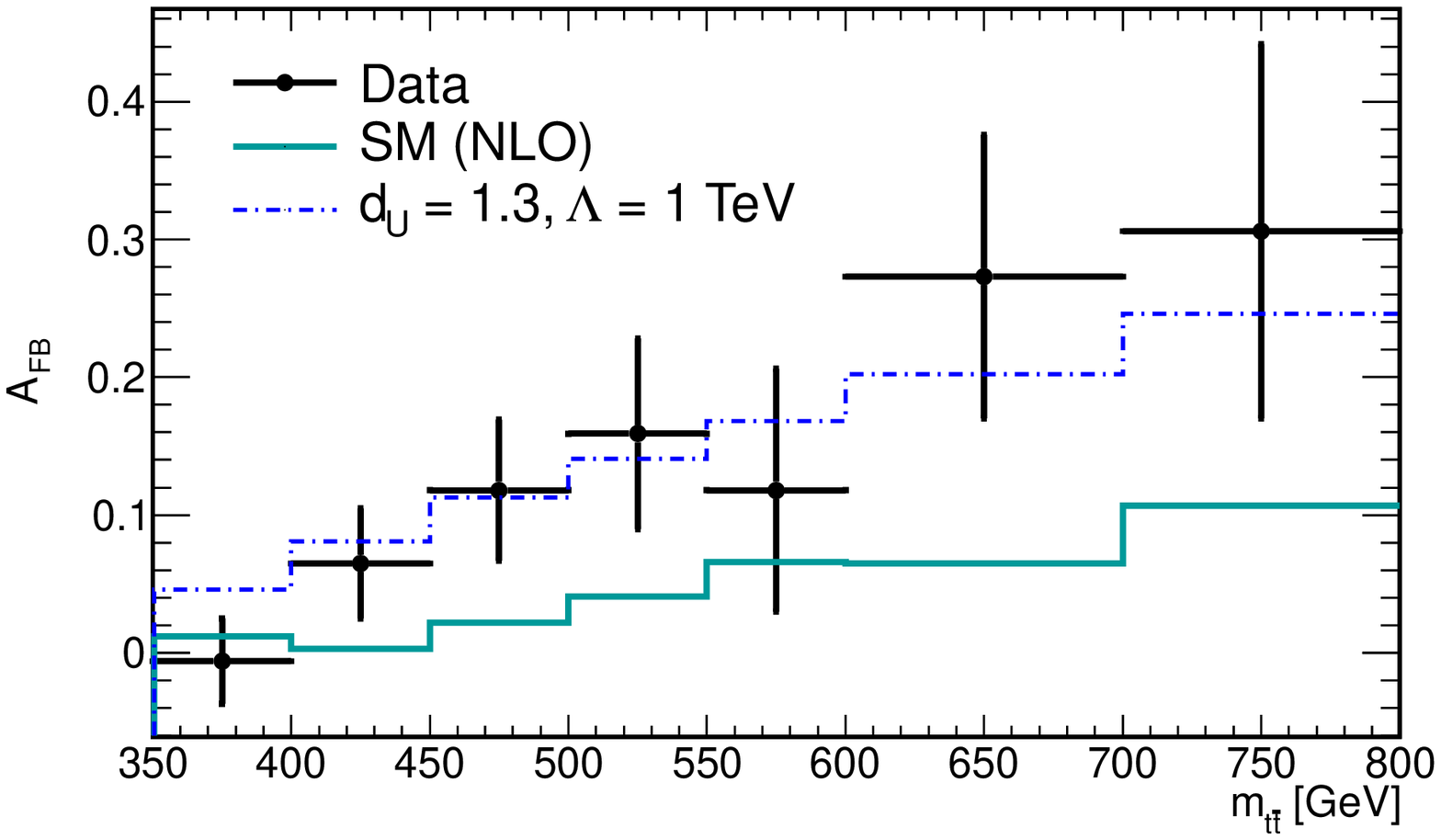}
  \caption{\small{Left: The $68\%$ C.L. region of $\Lambda$ (in GeV) in terms of $d_{\mathcal{U}}$ consistent with Tevatron measurement of
$A_{\text{FB}}$, $\sigma_{\text{TeV}}$, $\sigma_{\text{LHC}}$. The right side of the dashed green curve is
      consistent with the dijet cross section measured by CDF. Right: The Tevatron $A_{\text{FB}}$ versus top pair invariant mass.}}\label{combine}
\end{figure}

Now we combine the observables $A_{\text{FB}}$ (considering the
available measured values in all bins of $m_{t\bar{t}}$), $\sigma_{\text{LHC}}$, and
$\sigma_{\text{TeV}}$ into a global $\chi^{2}$ fit to
obtain $68\%$ C.L region.
The results of the global $\chi^{2}$ fit together with the constraints
arising from dijet resonance searches are presented in
Fig.\ref{combine} (left).
Unparticles can contribute to the production
of dijet at the Tevatron and LHC \cite{Agarwal}. We studied the
dijet production at parton level in unparticle model and compared the
results with the dijet invariant mass spectra measured by the CDF
experiment at the Tevatron \cite{dijet}. The allowed region is
depicted in Fig.\ref{combine} (left) in the right side of the green dashed curve.
We observe that the dijet constraints reduce the region where $A_{\text{FB}}$
could be generated according to the Tevatron
measurements. We note that when we move toward the large
values of $\Lambda$, the allowed area in the parameter space which can
produce the desired forward-backward asymmetry gets smaller. For any valid
value of $d_{\mathcal{U}}$, the dijet analysis excludes the region of
$\Lambda$ above 10 TeV.

The CDF experiment has measured the forward-backward asymmetry $A_{\text{FB}}$
in different $t\bar{t}$ invariant mass bins. In Fig.\ref{combine} (right), $A_{\text{FB}}$ is presented including data, the NLO SM
prediction, and the unparticle expectation with $d_{\mathcal{U}} = 1.3, \Lambda
= 1$ TeV. We  note that $d_{\mathcal{U}} = 1.3$ is the best fitted
point for $\Lambda
= 1$ TeV. Except for the first invariant mass bin ($m_{t\bar{t}} \in
(350,400)$) that unparticle has predicted larger forward backward asymmetry
than the experimental measurement, other bins show consistency with the
measurements. However, our results are compatible with the measurements within $1\sigma$.



\section{Conclusions}
New physics models that have been proposed to explain the observed Tevatron forward-backward asymmetry
are expected to affect the $t\bar{t}$ observables at the Tevatron and
LHC. Therefore, the new measurements
 are able to constrain the parameter space of the new models or discard the models. In this paper,
we have performed an analysis to address the observed forward backward
asymmetry of top at the Tevatron considering the color singlet vector
unparticles. We have examined the essential observables of the model
at the Tevatron and LHC including the total cross sections, the LHC
charge asymmetry, the $t\bar{t}$ invariant mass distribution and
dijet invariant mass spectra. In spite of the significant tension
between the reported forward backward asymmetry of top at the Tevatron
with other experimental measurements, we have found a small region in the
space of parameters of the color singlet vector
unparticle which can reproduce the $A_{\text{FB}}$ without being in
tight conflict with other $t\bar{t}$ measurements.
It has been shown that the data from dijet resonance searches reduces the parameter space
where the $A_{\text{FB}}$ can be generated according to the Tevatron observation. In particular, for any value of
$d_{\mathcal{U}}$, dijet data excludes unparticles with $\Lambda > 10$ TeV which have been compatible
with $t\bar{t}$ observables.

{\bf Note Added:} While this analysis was being completed, a related work appeared in \cite{Dahiya:2012ka}.

\end{document}